\documentstyle[pra,preprint,aps]{revtex}
\tightenlines
\begin{document}
\draft

\title{\bf POLARIZATION OF RADIATION IN MULTIPOLE JAYNES-CUMMINGS MODEL}

\author{Muhammet Ali Can and Alexander S. Shumovsky}

\address{Faculty of Science, Bilkent University, Bilkent, Ankara,
06533, Turkey}

\maketitle

\begin{abstract}
We discuss the spatial properties of quantum radiation emitted by
a multipole transition in a single atom. The qualitative
difference between the representations of plane and spherical
waves of photons is examined. In particular, the spatial
inhomogeneity of the zero-point oscillations of multipole field is
shown. We show that the vacuum noise of polarization is
concentrated in a certain vicinity of atoms where it strongly
exceeds the level predicted by the representation of the plane
waves. A new general polarization matrix is proposed. It is shown
that the polarization and its vacuum noise strongly depend on the
distance from the source.
\end{abstract}

\narrowtext

\newpage

\section{Introduction}

It is well known that the Jaynes-Cummings model \cite{1} plays an
important role in investigation of interaction between atoms and
quantum radiation field (e.g., see \cite{2,3,4,5}). The point is
that the model describes fairly well the physical processes in the
system and, at the same time, allows an exact solution.

In the usual formulation of the Jaynes-Cummings model
\cite{1,2,3,4,5}, the atom is considered as though it consists of
two or very few non-degenerated levels. In fact, the radiative
transitions in real atoms occur between the states with given
angular momentum $j \geq 1$ and its projection $m=-j, \cdots ,j$
(e.g., see \cite{6}). This means that even in the case of only two
levels, the degeneration with respect to the quantum number $m$
taking $(2j+1)$ different values should be taken into account. The
simplest example is provided by a dipole transition between the
states $|j=1,m=0, \pm 1 \rangle$ and $|j'=0,m'=0 \rangle$ when the
excited atomic state is a triply degenerate one (see figure 1).

Let us stress one more important difference. The radiation field
in conventional Jaynes-Cummings model is represented by the plane
waves of photons with given linear momentum and polarization. At
the same time, the multipole transitions in real atoms emit the
multipole photons represented by the quantized spherical waves
with given angular momentum and parity \cite{7,8}. Although there
is no principle difference between the plane and spherical waves
within the classical domain since both represent the complete
orthogonal sets of solutions of the homogeneous wave equation
\cite{9} and can be re-expanded with respect to each other, the
quantum counterparts of these two representations are
non-equivalent because they describe the physical quantities (the
linear and angular momenta respectively) which cannot be measured
at once.

Moreover, the multipole field is characterized by more quantum
degrees of freedom than the plane waves of photons. In fact, the
monochromatic pure $j$-pole multipole radiation of a given type
(either electric or magnetic) is specified by the $(2j+1)$
different values of the quantum number $m$. Since $j \geq 1$, the
total number of degrees of freedom here is not less than three. At
the same time, the monochromatic plane waves of photons are
described by only two different polarizations. In particular, the
increase of the number of degrees of freedom can lead to an
increase of the zero-point oscillations \cite{10}.

The multipole generalization of the Jaynes-Cummings model has been
discussed in \cite{11,12}. Let us stress that similar models have
been considered in different problems of interaction of quantum
light with matter (e.g., see \cite{13,14,15,16} and references
therein).

The main objective of this paper is to examine the quantum
properties of light emitted by a dipole atom at any distance from
the source, depending on the boundary conditions. The paper is
arranged as follows. In section 2 we review the properties of quantum
multipole field in comparison with these of plane waves of
photons. In section 3 we briefly discuss the multipole generalization
of the Jaynes-Cummings model. Then, in section 4 we consider the
polarization of multipole radiation and introduce a novel general
polarization matrix. This object permits us to take into account
the spatial anisotrophy of both the electric and magnetic fields
at once. In section 5 we examine the spatial properties of multipole
photons emitted by an atom in an ideal cavity as well as in empty
space. In particular, we show that the polarization properties of
quantum multipole radiation changes with distance from the atom.
In section 6 we briefly discuss the obtained results.

\section{Quantum multipole field}

Following \cite{7,8,17}, we list below some important formulas
describing the quantum multipole field. It is usually considered
in the so-called helicity basis \cite{17}
\begin{eqnarray}
{\vec \chi}_{\pm} = \mp \frac{{\vec e}_x \pm i{\vec
e}_y}{\sqrt{2}} , \quad \quad {\vec \chi}_0={\vec e}_z. \label{1}
\end{eqnarray}
It is clear that $\{ {\vec \chi}_{\mu} \}$ formally coincides with
the three eigenstates of spin $1$ of a photon. Since the
polarization is defined to be the spin state of photons \cite{18},
one can choose to interpret ${\vec \chi}_{\pm}$ as the unit vector
of circular polarization with either positive or negative
helicity, while ${\vec \chi}_0$ gives the linear polarization in
the $z$-direction. To within the sign at ${\vec \chi}_{\pm}$ the
helicity basis (1) coincides with the so-called polarization basis
usually used in optics \cite{19}. In the basis (1), the
positive-frequency part of the operator vector potential of
multipole field can be expanded as follows \cite{7,17}
\begin{eqnarray}
{\vec A}_{\lambda}({\vec r})= \sum_k \sum_{\mu =-1}^1
\sum_{j=1}^{\infty} \sum_{m=-j}^j (-1)^{\mu} {\vec \chi}_{- \mu}
V_{\lambda kjm \mu }({\vec r})a_{\lambda kjm}, \label{2}
\end{eqnarray}
where $a_{\cdots}$ is the photon annihilation operator which obeys
the following commutation relations
\begin{eqnarray}
[a_{\lambda kjm},a^+_{\lambda' k'j'm'}]= \delta_{\lambda \lambda'}
\delta_{kk'} \delta_{jj'} \delta_{mm'}. \nonumber
\end{eqnarray}
Here $k$ is the wave vector, $\lambda = E,M$ denotes the type of
radiation (parity), index $j \geq 1$ gives the angular momentum,
index $m= -j , \cdots ,j$, and $a_{\lambda kjm}$ is the
annihilation operator of the corresponding photon. The mode
functions $V_{\cdots}({\vec r})$ in (2) are represented in the
following way
\begin{eqnarray}
V_{Ekjm \mu} & = & \gamma_{Ekj} [ \sqrt{j} f_{j+1}(kr) \langle
1,j+1, \mu, m- \mu |jm \rangle Y_{j+1,m- \mu}( \theta \phi )
\nonumber \\ & - & \sqrt{j+1}f_{j-1}(kr) \langle 1,j-1, \mu ,m-
\mu |jm \rangle Y_{j-1,m- \mu}( \theta , \phi ) ], \nonumber
\\ V_{Mkjm \mu} & = & \gamma_{Mkj} f_j(kr) \langle
1,j, \mu ,m- \mu |jm \rangle Y_{j,m- \mu}( \theta , \phi )
\label{3}
\end{eqnarray}
for the electric and magnetic radiation respectively. Here
\begin{eqnarray}
\gamma_{Ekj} = \sqrt{\frac{2 \pi \hbar c}{k{\cal V}(2j+1)}}, \quad
\quad \gamma_{Mkj} = \sqrt{ \frac{2 \pi \hbar c}{k{\cal V}}},
\nonumber
\end{eqnarray}
are the normalization constants, ${\cal V}$ is the volume of
quantization, $\langle \cdots |jm \rangle$ denotes the
Clebsch-Gordon coefficient of vector addition of the spin and
orbital parts of the angular momentum of a multipole photon, and
$Y_{\ell m}$ is the spherical harmonics. The radial contribution
into the mode function (3) depends on the boundary conditions. In
the standard case of quantization in terms of standing spherical
waves in a spherical cavity \cite{7,17}, we have
\begin{eqnarray}
f_{\ell}(kr)=j_{\ell}(kr) \equiv \sqrt{\frac{\pi}{2kr}} J_{\ell
+1/2}(kr), \label{4}
\end{eqnarray}
where $j_{\ell}(x)$ is the spherical Bessel function.

The positive frequency parts of the operator field strengths obey
the following  relations
\begin{eqnarray}
{\vec E}_{Ekjm} & = & i \sum_{\mu ,k,j,m} k(-1)^{\mu}{\vec
\chi}_{- \mu} V_{Ekjm \mu}({\vec r})a_{Ekjm}, \nonumber \\ {\vec
E}_{Mkjm} & = & i \sum_{\mu ,k,j,m} k(-1)^{\mu}{\vec \chi}_{- \mu}
V_{Mkjm \mu}({\vec r})a_{Mkjm}, \nonumber \\ {\vec B}_{Ekjm} & = &
-i \sum_{\mu ,k,j,m} k(-1)^{\mu}{\vec \chi}_{- \mu} V_{Mkjm
\mu}({\vec r})a_{Ekjm}, \nonumber \\ {\vec B}_{Mkjm} & = & i
\sum_{\mu ,k,j,m} k(-1)^{\mu}{\vec \chi}_{- \mu} V_{Ekjm
\mu}({\vec r})a_{Mkjm}. \label{5}
\end{eqnarray}
It can be easily seen from (2) and (5) that the electric multipole
field always has longitudinal component of the electric field
strength in addition to the two transversal components, while it
is completely transversal with respect to magnetic induction. At
the same time, the magnetic multipole field has all three
components of magnetic induction and only two transversal
components of the electric field strength.

The position dependence of the mode functions (3) is not an
unusual fact. In reality, the mode functions of the plane waves
also depend on position:
\begin{eqnarray}
{\vec A}^{(plane)}= \sum_k \sqrt{\frac{2 \pi \hbar c}{kV}}
\sum_{\mu = \pm 1} (-1)^{\mu} {\vec \chi}_{- \mu}e^{i{\vec k}
\cdot {\vec r}}{\bf a}_{k \mu}. \label{6}
\end{eqnarray}
Here we choose the basis (1) with ${\vec \chi}_0={\vec k}/k$ and
${\bf a}_{\cdots}$ denotes the photon annihilation operator,
corresponding to the states with given linear momentum (direction
of propagation) and transversal polarization with either helicity.
The third projection of the photon spin is forbidden in this case
\cite{18}.

To complete the discussion of the quantum multipole field,
consider now the zero-point or vacuum oscillations. Since the
free-field Hamiltonian has the form \cite{7,8,17}
\begin{eqnarray}
H= \sum_k \hbar \omega_k \sum_{\lambda ,j,m} (a^+_{\lambda
kjm}a_{\lambda kjm}+1/2), \nonumber
\end{eqnarray}
for the energy of the vacuum state in the whole volume of
quantization we get
\begin{eqnarray}
H_{vac}= \sum_k \hbar \omega_k \sum_{\lambda ,j,m} 1/2= \sum_k
\hbar \omega_k \left( \sum_j (2j+1) \right). \label{7}
\end{eqnarray}
For comparison, we show here well known expressions valid in the
case of plane waves of photons (6):
\begin{eqnarray}
H^{(plane)}= \sum_{k, \mu} \hbar \omega_k (a^+_{k \mu}a_{k
\mu}+1/2), \nonumber \\ H^{(plane)}_{vac}= \sum_k \hbar \omega_k
\sum_{\mu = \pm 1} 1/2= \sum_k \hbar \omega_k. \label{8}
\end{eqnarray}
Due to the definition of $k$, both expressions (7) and (8) give an
infinite energy and, at first sight, cannot be compared with each
other. In fact, this infinity is inessential because of the
following reason. The contribution of zero-point oscillations can
be observed only via measurement which implies an averaging of
physical quantities over a finite "volume of detection" and
exposition time of detector \cite{20,21}. Such an averaging plays
a part of filtration leading to a selection of a certain finite
transmission frequency band. It is therefore seen from (7) and (8)
that, even if the filtration process leads to selection of the
electric dipole photons only, the ratio of contributions of the
zero-point oscillations of the multipole field and plane waves is
equal to $3/2$. From the physical point of view, this result is
caused by the more number of quantum degrees of freedom in the
case of multipole photons \cite{10}.

Much more interesting and important result can be obtained from
consideration of the spatial properties of the zero-point
oscillations. The energy density of the field is represented as
follows
\begin{eqnarray}
H({\vec r})= \frac{1}{16 \pi} [{\cal E}^2({\vec r})+{\cal
B}^2({\vec r})]= \frac{1}{16 \pi} \left\{ [{\vec E}^+({\vec
r})+{\vec E}({\vec r})]^2+[{\vec B}^+({\vec r})+{\vec B}({\vec
r})]^2 \right\} . \nonumber
\end{eqnarray}
Then, the zero-point contribution is
\begin{eqnarray}
\langle 0|H({\vec r})|0 \rangle = \frac{1}{16 \pi} \left\{ \langle
0|{\vec E} \cdot {\vec E}^+|0 \rangle + \langle 0|{\vec B} \cdot
{\vec B}^+|0 \rangle \right\} . \label{9}
\end{eqnarray}
Consider first the case of plane waves. In view of (6) and
symmetry relations $E_x=B_y, \quad E_y=-B_x$, we get
\begin{eqnarray}
\langle 0|H^{(plane)}({\vec r})|0 \rangle = \frac{1}{V} \sum_k
\hbar \omega_k. \label{10}
\end{eqnarray}
Thus, the zero-point oscillations of the energy density of the
plane waves of photons are homogeneous in the space of
quantization in spite of the position dependence of the mode
functions in (6).

In turn, employing the relations (5), we get
\begin{eqnarray}
\langle 0|H({\vec r})|0 \rangle = \frac{1}{8 \pi} \sum_k k^2
\sum_{\lambda ,j,m, \mu} |V_{\lambda kjm \mu}({\vec r})|^2.
\label{11}
\end{eqnarray}
In view of (3), we obtain
\begin{eqnarray}
\langle 0|H({\vec r})|0 \rangle = \frac{1}{4V} \sum_k \hbar
\omega_k \sum_{j,m, \mu} [ \frac{1}{2j+1} | \sqrt{j} f_{j+1}(kr)
\langle 1,j+1, \mu, m- \mu |jm \rangle Y_{j+1,m- \mu} ( \theta
\phi ) \nonumber \\ - \sqrt{j+1} f_{j-1}(kr) \langle 1,j-1, \mu
,m- \mu |jm \rangle Y_{j-1,m- \mu}( \theta , \phi )|^2 \nonumber
\\ + |f_j(kr) \langle 1,j, \mu ,m- \mu |jm \rangle Y_{j,m- \mu}( \theta
, \phi )|^2 ] . \label{12}
\end{eqnarray}
Taking into account that
\begin{eqnarray}
Y_{\ell ,m- \mu}( \theta , \phi ) \sim e^{i(m- \mu ) \phi},
\nonumber
\end{eqnarray}
it is  straightforward to show that the zero-point energy density
of the multipole field is independent of the angular variables
$\theta$ and $\phi$, while depends on the distance from the origin
(the singular point corresponding to the source location):
\begin{eqnarray}
\langle 0|H({\vec r})|0 \rangle = \langle 0|H(r)|0 \rangle .
\nonumber
\end{eqnarray}
Thus, the density of the zero-point oscillations of multipole
field has the spherical symmetry with respect to the singular
point (atom), while manifests the radial dependence. Taking into
account the radial dependence (4) and making use of the following
formula \cite{22}
\begin{eqnarray}
\sum_{\jmath =0}^{\infty} J^2_{\jmath +1/2}(z)= \frac{1}{\pi}
Si(2z) \equiv \frac{1}{\pi} \int_0^{2z} \frac{\sin t}{t} dt ,
\nonumber
\end{eqnarray}
it is easy to prove that the multipole zero-point oscillations
(12) vanish at far distance. Thus, the vacuum fluctuations of the
multipole field are concentrated near the atom.

Further employing the properties of spherical Bessel functions
shows that the principal contribution into (12) at $kr \rightarrow
0$ comes from the term with $j_0(kr)$, corresponding to the
electric dipole radiation. The radial dependence of (12) at fixed
$k$ and $j=1$ is shown in figure 2. It is seen that the zero-point
oscillations of the multipole field are concentrated in some
vicinity of the singular point where they strongly exceed the
level (10) predicted within the framework of the representation of
plane waves of photons. From the figure 2, the radius of the region
of concentration of the zero-point oscillations can be estimated
as follows:
\begin{eqnarray}
r_0 \sim \frac{2}{k} = \frac{\lambda_k}{\pi} \sim
\frac{\lambda_k}{3}, \label{13}
\end{eqnarray}
where $\lambda_k$ is the wavelength. Thus, the atom condenses the
zero-point oscillations in the near and intermediate zones. Let us
stress that in the number of modern experiments on engineered
entanglement in the systems of trapped Rydberg atoms, the
interatomic distances are of the same order \cite{23}.

The above result (13) should be considered as an estimation from
below because it corresponds to the first term in (12). Successive
taking into account of further terms leads to a certain shift of
$r_0$ into the intermediate zone.

\section{Multipole Jaynes-Cummings model}

The main aim of this section is to emphasize that the interaction
between the photons and electrons on atomic sub-levels with the
same $j$ and different $m$ is specified by a certain coupling
constant independent of $m$.

Consider a two-level atom with the electric dipole transition $j=1
\rightarrow j'=0$. The coupling constant of the atom-field
interaction can be found by calculating the matrix element
\cite{8,17}
\begin{eqnarray}
g=- \frac{e}{2m_ec} \langle 0,0|{\vec p} \cdot {\vec A}+{\vec A}
\cdot {\vec p}|1,m \rangle  = ik_0 \langle 0,0|{\vec d} \cdot
{\vec A}|1,m \rangle , \label{14}
\end{eqnarray}
obtained from the expression
\begin{eqnarray}
\frac{1}{2m_e} \left( {\vec p}- \frac{e}{c} {\vec A} \right)^2 ,
\nonumber
\end{eqnarray}
describing the interaction between the atomic electron with linear
momentum $\vec p$, charge $e$, and mass $m_e$ and radiation field
specified by the vector potential $\vec A$. Here ${\vec d}=e{\vec
r}$ is the dipole moment of the atomic transition with the
resonance frequency $\omega_0 =ck_0$. Assuming the central
symmetry of atomic field and taking into account the fact that the
spin state of an atom does not change under the electric dipole
transition, we can represent the atomic states in (14) as follows
\begin{eqnarray}
\langle (\vec r)|1,m \rangle = {\cal R}_e(kr)Y_{1m}( \theta , \phi
), \quad \quad\langle (\vec r) |0,0 \rangle = {\cal R}_g(kr)Y_{00}( \theta , 
\phi ), \nonumber
\end{eqnarray}
where ${\cal R}_{e,g}$ is the radial part of the atomic wave
function of either excited ($e$) or ground ($g$) state.

Expanding  the dipole moment $\vec d$ over the helicity basis (1),
substituting (2), and carrying out the calculations of integrals
in (14) over a small volume occupied by the atom, for the coupling
constant (14) we get
\begin{eqnarray}
g= \frac{k_0c}{\sqrt{kc}} D, \label{15}
\end{eqnarray}
where $D$ is the effective dipole factor which, by construction,
is independent of the quantum number $m$. Taking into account the
properties of the Clebsch-Gordon coefficients and spherical
harmonics, for the position-dependent mode function in (2) we get
\begin{eqnarray}
\lim_{kr \rightarrow 0} V_{Ek1m \mu} \sim \delta_{m \mu}.
\nonumber
\end{eqnarray}
This means that the electric dipole transition $|1,m \rangle
\rightarrow |0,0 \rangle$ at any given $m$ creates a photon with
helicity (polarization) $\mu =m=0, \pm 1$.

Finally, the Jaynes-Cummings Hamiltonian of the electric dipole
transition in the rotating-wave approximation \cite{24} takes the
form \cite{11,12}
\begin{eqnarray}
\hbar^{-1} H=H_0+H_{int}, \nonumber \\ H_0  =   \sum_{m=-1}^1 \{
\omega a^+_{m}a_{m}+ \omega_0 R_{mm} \} , \nonumber \\ H_{int} = g
\sum_{m=-1}^1 \{ R_{mg}a_{m}+a^+_{m}R_{gm} \} . \label{16}
\end{eqnarray}
To simplify the notation, hereafter we omit insignificant indices.
Here the atomic operators are defined as usually \cite{24} in
terms of the projections on the atomic states:
\begin{eqnarray}
R_{mg}=|1,m \rangle \langle 0,0|, \quad \quad R_{mm'}=|1,m \rangle
\langle 1,m'|. \nonumber
\end{eqnarray}
The Hamiltonian (16) describes the creation and absorption of the
single cavity-mode photons {\it at the atom location}. Everywhere
in the surrounding space, we have to take into account the spatial
dependence of the radiation field described by the vector
potential (2). Similar model can be constructed in the case of
magnetic dipole radiation as well as in the case of other
high-order atomic multipoles.

\section{Polarization of multipole radiation}

It is known that the polarization defines the direction of
oscillations of the field strengths. Within the classical picture
based on the consideration of plane waves, the polarization is
defined to be the measure of {\it transversal} anisotrophy of the
electric field strength \cite{19}. In turn, the quantum mechanics
interprets the polarization as given spin state of photons
\cite{18}. In the usual approach, the quantitative description of
polarization is based either on the Hermitian Polarization matrix
or on the equivalent set of real Stokes parameters. In the
standard case of plane waves, we get the $(2 \times 2)$
polarization matrix and four Stokes parameters \cite{19}, while
the description of multipole radiation requires for the $(3 \times
3)$ polarization matrix and nine Stokes parameters \cite{25,26}.
Moreover, the electric- and magnetic-type radiation fields are
usually described in terms of different polarization matrices,
taking into account the spatial anisotrophy \cite{27}.

Here we construct a more general novel object, describing in a
unique way the polarization properties of multipole radiation of
either type both classical and quantum as well as these of the
plane waves and other forms of electromagnetic radiation (e.g., of
the cylindrical waves).

In general, the field can be described in terms of the
field-strength tensor which can be chosen as follows \cite{28}
\begin{eqnarray}
F= \left( \begin{array}{cccc} 0 & E_x & E_y & E_z \\ -E_x & 0 &
-B_z & B_y \\ -E_y & B_z & 0 & -B_x \\ -E_z & -B_y & B_x & 0
\end{array} \right) \label{17}
\end{eqnarray}
It seems to be tempting to introduce the general quantitative
description of polarization using (17). Since the polarization is
specified by the intensities of different spatial components of
the radiation field and by the phase differences between these
components \cite{19}, it should be described in terms of bilinear
forms in the field strengths. The simplest bilinear form in the
field-strength tensor is
\begin{eqnarray}
R=F^+F, \label{18}
\end{eqnarray}
which differs from the energy-momentum tensor by a scalar. In some
sense it is similar to the Ricci tensor considered in the general
relativity \cite{29}. It is easily seen that (18) has the
following structure
\begin{eqnarray}
R= \left( \begin{array}{cc} W_E & {\vec S} \\ {\vec S}^+ & P
\end{array} \right)  \nonumber
\end{eqnarray}
where $W_E \equiv {\vec E}^+ \cdot {\vec E}$ is a scalar, ${\vec
S}$, apart from an unimportant factor, coincides with the Poynting
vector, and $P$ is the Hermitian $(3 \times 3)$ matrix of the form
\begin{eqnarray}
P=P_E+P_B. \label{19}
\end{eqnarray}
Here
\begin{eqnarray}
P_E= \left( \begin{array}{ccc} E^+_xE_x & E^+_xE_y & E^+_xE_z \\
E^+_yE_x & E^+_yE_y & E_y^+E_z \\ E^+_zE_x & E^+_zE_y & E^+_zE_z
\end{array} \right) \label{20}
\end{eqnarray}
and
\begin{eqnarray}
P_B= \left( \begin{array}{ccc} B^+_yB_y+B^+_zB_z & -B^+_yB_x &
-B^+_zB_x \\ -B^+_xB_y & B^+_xB_x+B^+_zB_z & -B^+_zB_y \\
-B^+_xB_z & -B^+_yB_z & B^+_xB_x+B^+_yB_y \end{array} \right)
\label{21}
\end{eqnarray}
We note here that (20) has been proposed in \cite{25} in order to
describe the spatial anisotrophy of the electric dipole radiation,
while (21) is similar to the object has been discussed in
\cite{26,27}.

We choose to interpret (19) as the general polarization matrix,
while the terms (20) and (21) give the electric and magnetic field
contributions respectively.

To justify this statement, consider first the case of plane waves
propagating in the $z$-direction when $E_z=B_z=0$ and $B_x=-E_y,
\quad B_y=E_x$. Then, the matrix(20) takes the form
\begin{eqnarray}
P_E^{(plane)}= \left( \begin{array}{ccc} E^+_xE_x & E^+_xE_y & 0
\\ E^+_yE_x & E^+_yE_y & 0 \\ 0 & 0 & 0 \end{array} \right)
\label{22}
\end{eqnarray}
It is seen that the non-zero submatrix in (22) coincides with
conventional $(2 \times 2)$ polarization matrix of plane waves
\cite{19,21}. In turn, (21) takes the form
\begin{eqnarray}
P_B= \left( \begin{array}{ccc} E^+_xE_x & E^+_xE_y & 0
\\ E^+_yE_x & E^+_yE_y & 0 \\ 0 & 0 & {\vec E}^+ \cdot {\vec E}
\end{array} \right)
\label{23}
\end{eqnarray}
where corresponding $(2 \times 2)$ submatrix in the top left
corner coincides with (22). Thus, the general polarization matrix
(19) describes the polarization of plane waves adequately.

Consider now the multipole radiation. In the case of electric-type
radiation when $B_z=0$ everywhere, the matrix $P_E$ has the
general form (20), while the magnetic polarization matrix (21) is
reduced to
\begin{eqnarray}
P_B= \left( \begin{array}{ccc} B^+_yB_y & -B^+_yB_x & 0 \\
-B^+_xB_y & B^+_xB_x & 0 \\ 0 & 0 & {\vec B}^+ \cdot {\vec B}
\end{array} \right)
\label{24}
\end{eqnarray}
The polarization of magnetic-type radiation is described by (20)
with $E_z=0$ which coincides with (22) and by the general form
(21) with $B_z \neq 0$. In this case, (21) coincides, within the
transposition of lines and columns, with the polarization matrix
considered in \cite{27}.

It is natural that the general polarization matrix (19) reflects
the three-dimensional structure of the radiation field. The
diagonal terms in (20) and (21) give the radiation intensities.
Their angular and radial dependence corresponds to the radiation
patterns of the multipole field. The off-diagonal terms give the
phase information as in the case of plane waves \cite{19}. In
contrast to the standard case of plane waves, there are the two
independent phase differences $\Delta_{ij} \equiv \arg E_i- \arg
E_j$ instead of only one phase difference because
\begin{eqnarray}
\Delta_{xy} +\Delta_{yz} + \Delta_{zx} =0. \nonumber
\end{eqnarray}
Since ${\vec E}({\vec r}) \cdot {\vec B}({\vec r})=0$ at any
point, the magnetic part (21) of the general polarization matrix
(19) contains the same phase differences as (20).

Similar expressions for the polarization matrix (19) can also be
obtained in the helicity basis (1). For example, the matrix (20)
takes the form
\begin{eqnarray}
P_E= \left( \begin{array}{ccc} E^+_+E_+ & -E^+_+E_0 & E^+_+E_- \\
-E^+_0E_+ & E^+_0E_0 & -E^+_0E_- \\ E^+_-E_+ & -E^+_-E_0 &
E^+_-E_- \end{array} \right) \label{25}
\end{eqnarray}
The quantum counterpart of (19) can be obtained by formal
substitution of the operators instead of the classical field
strengths (compare with \cite{30}). Averaging of the corresponding
operator matrix over a given state of the radiation field gives
the polarization matrix. By construction, the operator matrices
(19)-(24) correspond to the normal ordering in the creation and
annihilation operators:
\begin{eqnarray}
P=P(a^+a). \nonumber
\end{eqnarray}
In addition, one can define the anti-normal polarization matrix
\begin{eqnarray}
P^{(an)}=P(aa^+) \nonumber
\end{eqnarray}
by a simple change of order of product of the field strengths in
all elements of the matrices (19)-(24). It is then clear that the
matrix
\begin{eqnarray}
P^{(an)}-P=P([a,a^+])= \langle 0|P^{(an)}|0 \rangle \equiv P_{vac}
\label{26}
\end{eqnarray}
determines the zero-point (vacuum) contribution into the
polarization. Following the ideas and results of section 2, it is a
straightforward matter to show that the vacuum polarization of
plane waves of photons is uniform in the space, while the
multipole vacuum polarization concentrates near the atom and
exceeds the level predicted by the representation of plane waves.

\section{Polarization of a single-atom radiation}

It was shown in section 3 that an atomic electric dipole transition
with given $m$ emits the photon with given polarization $\mu =m$.
We now show that the polarization changes with the distance from
the source. In other words, the polarization is not a global
property of the field, while changes from point to point, at least
in the case of multipole radiation.

Assume, for example, that the atom emits the electric dipole
photon with $m=+1$, i.e. circularly polarized with positive
helicity. Consider the polar direction $( \theta = 0)$ ,
corresponding to the maximum of the radiation pattern in this case
\cite{9}. Then, the matrix (25) averaged over the photon state
$|1_+ \rangle$ takes the form
\begin{eqnarray}
{\overline P}_E= \frac{\hbar \omega}{3V} \left(
\begin{array}{ccc} [ \frac{1}{2} j_{2}(kr)-j_{0}(kr)]^2 & 0 &
0 \\ 0 & 0 & 0 \\ 0 & 0 & 0 \end{array} \right) \nonumber
\end{eqnarray}
Thus, there is only one polarization $\mu =+1$ in the polar
direction. In the less probable case of the equatorial direction
$( \theta = \pi /2)$, from (25) we get
\begin{eqnarray}
{\overline P}_E= \frac{\hbar \omega}{3V} \left( \begin{array}{ccc}
[ \frac{1}{4} j_{2}(kr)+j_{0}(kr)]^2 & 0 & \frac{-3}{4} j_{2}(kr)
[ \frac{1}{4} j_{2}(kr)+j_{0}(kr)]e^{2i \phi} \\ 0 & 0 & 0
\\ \frac{-3}{4} j_{2}(kr) [ \frac{1}{4}
j_{2}(kr)+j_{0}(kr)]e^{-2i \phi} & 0 & \frac{9}{16} [j_{2}(kr)]^2
\end{array} \right) \label{27}
\end{eqnarray}
so that there are the two circularly polarized components with
opposite helicities. Comparison of intensities of the two
components shows that the positive helicity dominates at short
distances ($kr \leq 3$), while both components contribute equally
at far distances ($kr \gg 1$) (see figure 3). It is clear that any
deviation from the polar direction leads to creation of
polarizations additional to $\mu =+1$. Thus, the polarization o
radiation under consideration strongly depends on the direction
and distance from the source. Similar picture can be obtained for
polarization of photons with $m=-1$ and $m=0$.

The above results were obtained in the case of standing waves in
an ideal spherical cavity when the radial dependence of the mode
functions (3) is specified by equation (4). In this case, the radiation
field is subjected to the Rabi oscillations which can be described
through the use of the steady-state time dependent wave function
for the system with Hamiltonian (16):
\begin{eqnarray}
| \psi (t) \rangle =e^{-iHt}|e_m;0 \rangle = \frac{1}{2}
\sum_{\ell = \pm 1} e^{-i \ell gt}(|e_m;0 \rangle + \ell |g;1_m
\rangle ),  \label{28}
\end{eqnarray}
where we choose the initial state as the vacuum state of the
cavity field and excited state of the atom is specified by given
$m$. Then, the elements of the polarization matrix (19) obtained
by averaging or corresponding operator matrix over the state (28)
should be multiplied by an additional factor of $(1- \cos 2gt)$,
describing the steady-state time dependence of polarization.

Consider now the radiation by a dipole atom in empty space. Then,
the Hamiltonian (16) should be generalized as follows
\begin{eqnarray}
\hbar^{-1} H=H_0+H_{int}, \nonumber \\ H_0  =   \sum_{m=-1}^1 \{
\sum_k \omega_k a^+_{km}a_{km}+ \omega_0 R_{mm} \} , \nonumber \\
H_{int} = \sum_k g_k \sum_{m=-1}^1 \{ R_{mg}a_{km}+a^+_{km}R_{gm}
\} , \label{29}
\end{eqnarray}
to take into account the $k$-dependence of the radiation field.
Let us again choose the initial state as the vacuum state of
photons and excited atomic state with given $m$
\begin{eqnarray}
| \psi_0 \rangle =|e_m \rangle \otimes [ \bigotimes_k |0_k \rangle
]. \label{30}
\end{eqnarray}
It is then clear that the radiation field is represented by the
outgoing spherical waves of photons which stipulates the choice of
the radial dependence in (3) in terms of the spherical Hankel
function of the first kind
\begin{eqnarray}
f_{\ell}(kr)=h^{(1)}_{\ell}(kr)=j_{\ell}(kr)+i(-1)^{\ell+1}j_{-
\ell -1}(kr) \label{31}
\end{eqnarray}
instead of (4) \cite{31}. This choice assumes that the atom
occupies a small but finite spherical volume of radius $r_a$ at
the origin to avoid the divergence at $kr \rightarrow 0$.

The elements of the polarization matrix (25) of the electric
dipole radiation in the equatorial direction take the form
\begin{eqnarray}
E^+_+E_+ & = & \frac{\hbar \omega}{3V} \left\{ [ \Gamma_+ (kr)
]^2+[ \Gamma_- (kr) ]^2 \right\} , \nonumber \\ E_-^+E_- & = &
\frac{ \hbar \omega}{3V} \frac{9}{16} \left\{
[j_{2}(kr)]^2+[j_{-3}(kr)]^2 \right\} , \nonumber \\ E^+_+E_- & =
& \frac{-\hbar \omega}{4V} \{ [ j_2(kr)\Gamma_+ (kr)+
j_{-3}(kr)\Gamma_- (kr)]^2 \nonumber \\
&+&[j_{2}(kr)\Gamma_-(kr)-j_{-3}(kr)\Gamma_+(kr)]^2\}^{1/2} e^{i
\varphi} , \label{32}
\end{eqnarray}
where
\begin{eqnarray}
\Gamma_{+} (kr) & \equiv & \frac{1}{4} j_{2}(kr)+j_{0}(kr)
\nonumber
\\ \Gamma_{-} (kr) & \equiv & \frac{1}{4} j_{-3}(kr)+j_{-1}(kr) \nonumber
\end{eqnarray}
and
\begin{eqnarray}
\varphi \equiv 2 \phi + \tan^{-1} \left( \frac{\Gamma_-
(kr)j_{2}(kr)- \Gamma_+ (kr)j_{-3}(kr)}{\Gamma_+ (kr)j_{2}(kr)+
\Gamma_- (kr)j_{-3}(kr)} \right) . \nonumber
\end{eqnarray}
All elements containing $E_0$ are equal to zero.  Unlike the case
of radiation in an ideal cavity (27), the phase difference
$\varphi$ between the components with opposite helicity depends
here on the distance from the source. Other cases can be examined
in the same way.

For the polarization matrix (25) in the polar direction, we again
get only one non-zero element
\begin{eqnarray}
E^+_+E_+= \frac{\hbar \omega}{3V} \left\{ [ \Xi_+ (kr)]^2+[ \Xi_-
(kr)]^2 \right\}, \nonumber
\end{eqnarray}
where
\begin{eqnarray}
\Xi_{+} (kr) & \equiv & \frac{1}{2} j_{2}(kr)-j_{0}(kr) \nonumber
\\ \Xi_{-} (kr) & \equiv & \frac{1}{2} j_{-3}(kr)-j_{-1}(kr)
\nonumber
\end{eqnarray}
The time evolution can be described with the aid of approach
proposed in \cite{32}. Then, the matrix elements of the
polarization matrix should be multiplied by the following factor
\begin{eqnarray}
1-2e^{- \eta t/2} \cos [( \omega_k - \omega_0 - \Delta \omega )t]
+e^{- \eta t}, \nonumber
\end{eqnarray}
where
\begin{eqnarray}
\eta = \frac{\pi}{2} \sum_k g^2_k \delta ( \omega_k - \omega_0 ),
\quad \quad \Delta \omega = \frac{{\cal P}}{\omega_0 - \omega_k}
\nonumber
\end{eqnarray}
and $\cal P$ denotes the principle value of corresponding
integral. Again, any deviation from the polar direction changes
the picture of polarization at intermediate and far distances with
respect to that at the atom location.

\section{Conclusion}

Let us briefly discuss the obtained results. In this paper we have
concentrated on the description of the spatial properties of
multipole radiation by a single atom. The consideration is based
on sequential use of the representation of multipole photons
corresponding to the radiation of real atoms. It is shown that the
zero-point oscillations of the multipole field are concentrated
near the atom, while exceed the level predicted by the model of
plane waves of photons everywhere. In a certain neighborhood of
the atom, the effect is strong enough. Although the effect can be
observed at the relatively short distances, it seems to be
important for the near- and intermediate-field quantum optics as
well as for the experiments with trapped Rydberg atoms when the
typical interatomic distances are of the same order \cite{23}. In
particular, it can be important for engineered entanglement in the
system of two atoms in a cavity proposed in \cite{33}, for
experiments with single-atom laser as well as for the estimation
of Casimir effect in atomic systems.

Let us note in this connection that possible influence of an atom
on the electromagnetic vacuum state in the absence of radiation
has been discussed in quantum electrodynamics for a long time
(e.g., see \cite{34,35}). The new element here is the spatial
inhomogeneity of the vacuum noise. Unlike the effects discussed in
\cite{34,35}, the specific distance dependence of the zero-point
oscillations in the presence of atom has the geometrical nature
and is independent of the atom-field interaction.

To describe the polarization of multipole field, we proposed in
section 4 a new definition of the polarization matrix based on the
bilinear form in the field-strength tensor. The generalized
polarization matrix (19) is additive with respect to the
contributions coming from the electric field and magnetic
induction. It reflects the three-dimensional nature of
polarization connected with the three possible states of spin of
photon. In special case of plane waves, when the third spin state
is forbidden, it reduces to the conventional $(2 \times 2)$
Hermitian polarization matrix. In the case of multipole radiation,
it combines together the objects considered earlier in
\cite{25,26,27}.

By construction, the generalized operator polarization matrix is a
local object in spite of the global nature of the photon operators
of creation and annihilation. The spatial properties of
polarization are caused by the mode functions and changes with
distance and direction from the source. For example, the electric
dipole radiation from the excited atomic level with $m=+1$,
corresponding to the creation of a circularly polarized photon
with positive helicity, is transformed, at far distances, into the
radiation with both helicities. In the case of radiation in empty
space, the phase difference between the modes with opposite
helicities is also a function of distance and direction from the
source.

Both the distance and direction dependence of the polarization
seems to be very important. In fact, any real measurement of
intensity assumes the finite aperture of a detecting device
\cite{20,21}. This means that, in the case of radiation by the
atomic transition $\j=1,m=+1 \rangle \rightarrow |j'=0,m'=0
\rangle$ considered in section 5, the precision of measurement of the
polarization $\mu =+1$ in the polar direction would be influenced
by the zero-point oscillations of all three polarizations.

Let us stress that locality of polarization discussed in this
paper can also be interpreted in terms of the photon localization.
It is well known that, while the photon operators of creation and
annihilation are defined in the whole space at once, the notion of
photon localization is not deprived of physical meaning
\cite{21,34}. The photodetection process provides a quite certain
example of the photon localization \cite{20}. There are also known
attempts to interpret the photon localization as the specific
fall-off of the photon energy density \cite{37}. It is also clear
that emission and absorption of radiation by atoms can also be
interpreted as the photon localization \cite{38}. Our results
based on consideration of the representation of multipole photons
shows the rapid fall-off of the density of zero-point oscillations
and of the vacuum noise of polarization with the distance from
atom as well as the spatial dependence of a certain
characteristics of the radiation field (e.g., polarization).

One of the authors (A.S.Sh.) would like to thank Professor J.H.
Eberly, Prof. A.A. Klyachko, Professor V.I. Rupasov, and Professor
A. Vourdas for many fruitful discussions.

\begin{figure}
\caption{Energy diagram of triple degenerated excited and ground
states of a dipole transition $j=1 \leftrightarrow j'=0$.}
\end{figure}

\begin{figure}
\caption{Zero-point oscillations of the energy density versus
dimensionless distance $kr$ at fixed $k$ and $j=1$.}
\end{figure}

\begin{figure}
\caption{Distance dependence of intensity of the multipole
radiation generated by the atomic transition $|j=1,m=+1 \rangle
\rightarrow |j'=0,m'=0 \rangle$ in equatorial direction.}
\end{figure}

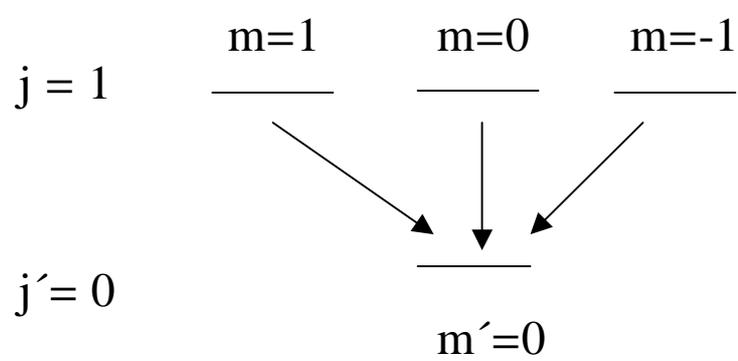
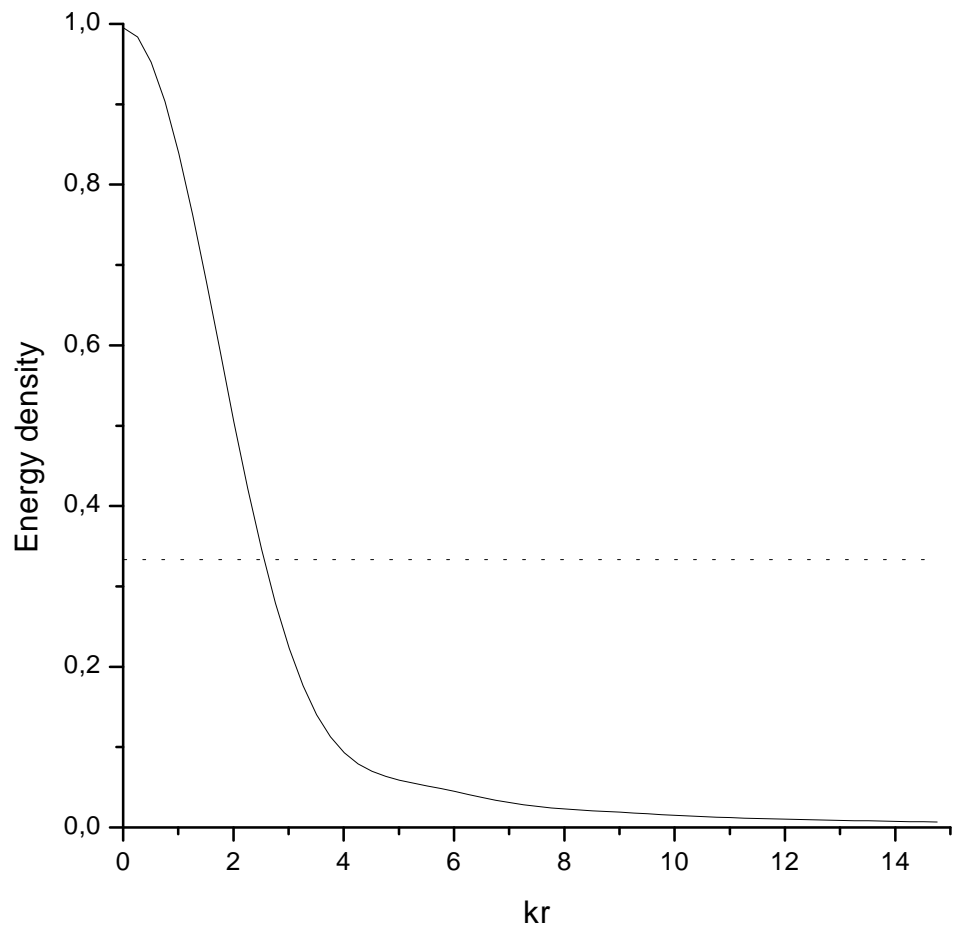
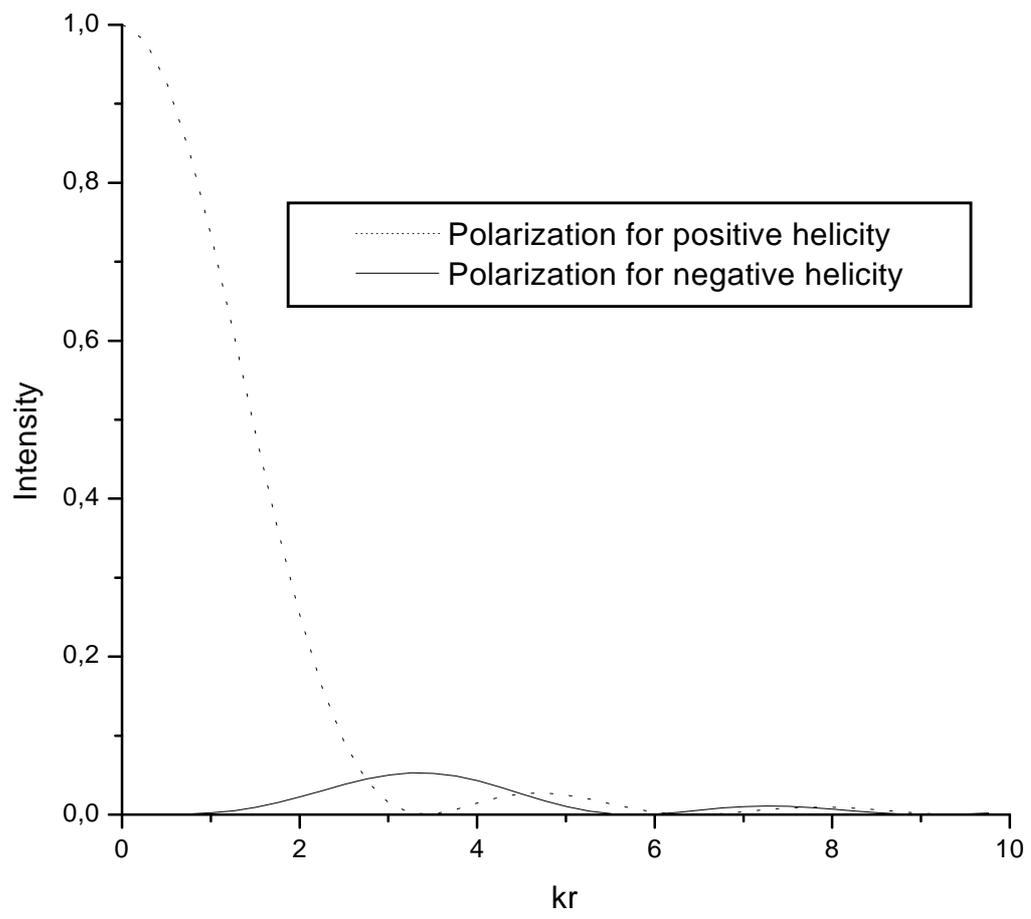
\end{document}